\documentclass[11pt,a4paper]{article}
\pdfoutput=1

\usepackage{amsmath,amsfonts,amsbsy,amssymb,array,accents}
\usepackage{enumerate,array,latexsym,graphicx,mathrsfs,verbatim,psfrag}
\usepackage{bm} 
\usepackage{booktabs} 
\usepackage[usenames]{color}
\usepackage{setspace}
\usepackage[english]{babel}
\usepackage{fancybox}
\usepackage{datetime}
\usepackage[nosort]{cite}
\usepackage{chngpage} 
\usepackage{textcomp,gensymb} 
\usepackage{wasysym}
\usepackage{psfrag}
\usepackage{epstopdf}  

\usepackage{float}
\usepackage{tikz}
\usetikzlibrary{decorations.text}
\usepackage{enumitem}
\setlist{itemsep=0pt}

\usepackage{braket}

\usepackage{fancyhdr}
\usepackage{datetime}

\usepackage{mciteplus}

\usepackage[colorlinks=true,      linkcolor=darkblue,      urlcolor=darkblue,      
            filecolor=darkblue,      citecolor=darkblue,       pdfstartview=FitH,     
						pdfpagemode=UseNone,      bookmarksopen=true]{hyperref}  

\usepackage[all]{hypcap}     

\newcommand{\captionfonts}{\small}

\makeatletter  
\long\def\@makecaption#1#2{%
  \vskip\abovecaptionskip
  \sbox\@tempboxa{{\captionfonts #1: #2}}%
 \ifdim \wd\@tempboxa >\hsize
    {\captionfonts #1: #2\par}
  \else
    \hbox to\hsize{\hfil\box\@tempboxa\hfil}%
  \fi
  \vskip\belowcaptionskip}
\makeatother   

\DeclareMathSymbol{\medhatsym}{\mathord}{largesymbols}{"62} 

\DeclareMathSymbol{\medtildesym}{\mathord}{largesymbols}{"65}

\newcommand{\comm}[1]{} 

\renewcommand{\arraystretch}{1.2}
\def\IC{\mathbb{C}}

\def\IP{\mathbb{P}}

\def\IR{\mathbb{R}}

\def\IZ{\mathbb{Z}}

\numberwithin{equation}{section} 

\makeatletter
\g@addto@macro\bfseries{\boldmath}
\makeatother

\definecolor{cardinal}{rgb}{0.6,0,0}
\definecolor{darkgreen}{rgb}{0,0.4,0}
\definecolor{purple}{rgb}{0.5, 0, 0.5}
\definecolor{golden}{rgb}{0.92, 0.7, 0}
\definecolor{midnight}{rgb}{0, 0, 0.5}
\definecolor{darkblue}{rgb}{0, 0, 0.7}

\def\IC{\mathbb{C}}
\def\Neql#1{{\cal N}\!=\!{#1}}
\def\coeff#1#2{\relax{\textstyle {#1 \over #2}}\displaystyle}

\def\IR{\mathbb{R}}

\def\IZ{\mathbb{Z}}

\def\cJ{{\cal J}}

\def\cN{{\cal N}}

\def\cP{{\cal P}}

\def\eql{~=~}

\def\so{\overset{_{\phantom{.}\circ}}{s}{}}
\def\go{\overset{_{\phantom{.}\circ}}{g}{}}

\def\Ro{\overset{_{\phantom{.}\circ}}{R}{}}

\begin{document}

\phantom{AAA}
\vspace{-10mm}

\begin{flushright}
%
%
\end{flushright}

\vspace{18mm}

\begin{center}
\begin{adjustwidth}{-5mm}{-5mm} 

\centerline{\huge \bf Brane-Jet Instabilities}


\end{adjustwidth}

\bigskip

\vspace{8mm}

{\bf {Iosif Bena$^1$,~Krzysztof Pilch$^{2}$   and  Nicholas P. Warner$^{1,2,3}$}}

\vspace{5mm}

$^1$Institut de Physique Th\'eorique, \\
Universit\'e Paris Saclay, CEA, CNRS,\\
Orme des Merisiers, Gif sur Yvette, 91191 CEDEX, France \\[12 pt]
\centerline{$^2$Department of Physics and Astronomy}
\centerline{and $^3$Department of Mathematics,}
\centerline{University of Southern California,} 
\centerline{Los Angeles, CA 90089, USA}

\vspace{6mm} 
{\footnotesize\upshape\ttfamily  iosif.bena @ ipht.fr, ~pilch @ usc.edu, ~warner @ usc.edu} \\

\vspace{15mm}
 
\textsc{Abstract}

\end{center}
\begin{adjustwidth}{6mm}{6mm} 
 
\vspace{-4mm}
\noindent

\noindent 
With one exception, all known non-supersymmetric  AdS$_4$ and AdS$_5$ vacua of gauged maximal supergravities that descend from string and M theory have been shown to have modes with mass below the BF bound. The remaining non-supersymmetric AdS solution is perturbatively stable within gauged maximal supergravity, and hence appears to contradict recent conjectures about the AdS stability based on the weak gravity conjecture.  We show that this solution is actually unstable by exhibiting a new decay channel, which is only visible when the solution is  uplifted to eleven dimensions. In particular,  M2 brane probes at generic locations inside the internal manifold are attracted to the Poincar\'e horizon, only to be expelled as ``brane jets'' along certain directions of the internal manifold. Such instabilities can arise in any non-supersymmetric AdS vacuum in any dimension.   When a brane-jet instability is present,   the force that expels the branes is the same as the force felt by a probe brane whose mass is less than its charge.

\bigskip

\end{adjustwidth}

\thispagestyle{empty}
\newpage


\baselineskip=13pt
\parskip=2.5pt

\setcounter{tocdepth}{2}
\tableofcontents

\baselineskip=15pt
\parskip=3pt


\section{Introduction} 
\label{sec:Intro}

The stability of an AdS vacuum is a subtle issue. It has been known for a very long time that potentials that give rise to ``negative masses,'' can have perturbatively stable vacua as long as those negative masses lie above the Breitenlohner-Freedman (BF) bound \cite{Breitenlohner:1982bm,Breitenlohner:1982jf}.  Moreover,  for supersymmetric AdS solutions one can establish positive mass theorems that rule out non-perturtive and tunneling  instabilities\footnote{See, for example, \cite{Gibbons:1983aq}.}. 

The interesting issue, therefore, is the stability of AdS vacua that break supersymmetry.  If there are scalars that violate the BF bound, these vacua are perturbatively unstable.  Furthermore, from the point of view of holography, these AdS vacua appear disastrous: the operators dual to the fields whose mass is below the BF bound have complex dimension, and hence the dual CFTs are not unitary. Thus, the only top-down examples of non-supersymmetric holography can come from BF stable, non-supersymmetric AdS vacua in the supergravities that descend from string theory \cite{Bobev:2011rv}.

Unfortunately, there is a startling lack of such vacua. This observation was advanced as evidence for the argument, based on the Weak Gravity Conjecture (WGC)  \cite{ArkaniHamed:2006dz}, that all top-down non-supersymmetric AdS vacua will somehow be unstable \cite{Ooguri:2016pdq}. The only fly in the ointment is the  $SO(3) \times SO(3)$ invariant vacuum  \cite{Warner:1983du,Warner:1983vz}, which is perturbatively (BF) stable in gauged $\cN =8$ supergravity in four dimensions.  

The scalar potentials of gauged maximal supergravity in four and five dimensions have now been extensively studied, and the last few years have seen the discovery of hundreds of non-supersymmetric AdS vacua and a handful of new supersymmetric ones.  The complete set of known  AdS$_4$ vacua may be found in   \cite{Warner:1983du,Warner:1983vz,Fischbacher:2009cj,Fischbacher:2010ec,Fischbacher:2011jx,Borghese:2013dja,Comsa:2019rcz,Bobev:2019dik} and their (in)stability is discussed in \cite{Gibbons:1983aq,Nicolai:1985hs,Bobev:2010ib,Fischbacher:2010ec,Fischbacher:2011jx,Borghese:2013dja,Comsa:2019rcz}.  The known   AdS$_5$ vacua may be found in \cite{Khavaev:1998fb,Krishnan:2020sfg,Bobev:2020ttg} and  their (in)stability is discussed in \cite{Freedman:1999gp,Freedman:1999dz,Pilch:1999misc,Distler:1999tr,Girardello:1998pd,Bobev:2020ttg}. All the new non-supersymmetric vacua violate the BF bound  and this represents further empirical support for the instability of non-supersymmetric AdS vacua \cite{Ooguri:2016pdq}.   It also makes the $SO(3) \times SO(3)$ invariant vacuum all the more unusual.

This solution was, in fact, the first non-trivial vacuum found in a gauged maximal supergravity \cite{Warner:1983du}, and its perturbative stability in  gauged maximal supergravity was established in \cite{Fischbacher:2010ec}. Furthermore, upon embedding this vacuum in gauged $\cN = 5$ supergravity one can prove a positive mass theorem  \cite{Boucher:1984yx} which might well be extensible to the full gauged $\cN =8$ theory. Thus it was believed that this vacuum was likely to prove stable.

The purpose of this paper is to highlight a new type of instability of AdS vacua that are created by branes and supported by fluxes:  {\it the Brane-Jet Instability}.  We will show how this destabilizes the non-supersymmetric, $SO(3) \times SO(3)$ invariant, AdS$_4$ vacuum of \cite{Warner:1983du} and that many BF-unstable vacua also have brane-jet instabilities.  In M theory, this instability is detected by M2 brane probes that lie parallel to the boundary of the  Poincar\'e patch  in AdS$_4$ and is only visible if one uplifts the gauged supergravity vacuum to M theory.   The instability  emerges through the dependence of the solution on the internal manifold, and specifically through the warp factor that plays an essential role in the uplift.    

Recall that, if $\widehat{ds}{}^2$ is the metric of a $d$-dimensional theory that appears as a factor in a higher-dimensional metric, then the full uplifted metric has the form 
\begin{equation}
ds^2 ~=~ \Delta^{-\frac{2}{d-2}} \, \widehat{ds}{}^2 ~+~ds_\text{int}^2 \,,
\label{upliftmet}
\end{equation}
where $ds_\text{int}^2$ is the metric on the internal manifold and $\Delta^2$ is the determinant of the internal metric.   The warp factor is required for Einstein gravity in the metric, $ds^2$, to reduce to Einstein gravity on  $\widehat{ds}{}^2$.  It is also essential to solving the equations of motion in the higher-dimensional theory.   As a result of this factor, the induced metric that enters in the DBI action of an M2 or D3 brane parallel to the boundary of the Poincar\'e patch depends on their location in the internal dimensions.  It is this dependence on the internal dimensions that can lead to the instability.  Using brane probes to test  stability is a standard technique (see for example,  \cite{Johnson:2000ic, Corrado:2001nv,  Gaiotto:2009mv});  the novelty of brane jets lies in mediation of instabilities through non-trivial warp factors.

When the vacuum has eight supercharges ($\Neql 2$), the M2 branes that are localized on a particular six-dimensional, transverse slice (corresponding to a Coulomb branch of the dual CFT)  feel no force  \cite{Corrado:2001nv}, and are attracted along other directions. However, as we will show in this paper, this supersymmetric background is very much the exception.  There are no such flat directions when the supersymmetry is reduced to $\Neql 1$: for such vacua the brane potential is always attractive, reflecting stability that is a consequence of supersymmetry.  For many of the non-supersymmetric AdS$_4$ vacua, including the $SO(3) \times SO(3)$ invariant one, we find that the M2 branes can be expelled  from the Poincar\'e horizon, along the direction of a ``jet'' that can be  sharply localized in the seven-dimensional internal space.

We will also show that, in a large number of non-supersymmetric examples,  if one averages over the internal space, the net attraction is always positive.  Thus smearing the probe or using the AdS effective Lagrangian that comes from averaging the metric (\ref{upliftmet}) over the internal space will generically wipe out the instability.

The brane-jet instability is different from the instability that is suggested by the na\"ive brane probe analysis that underlies the discussion in \cite{Maldacena:1998uz, Ooguri:2016pdq}. That analysis looks at the behavior of a hypothetical ``effective'' p-brane whose mass and charge are different and which is supposed to exist in some low-energy  theory in the AdS background.  While this might seem correct, the problem is that such an effective would-be probe brane does not have an obvious relation to the branes of string and M theory. 

Returning to the uplift of the supergravity solutions, it is important to note that  there could also be a different warp factor in front of the electric components of the flux along the brane directions. However, for AdS$_d$ backgrounds, in which the electric component of the field strength is proportional to the volume form of AdS$_d$, the Bianchi identities imply that the electric flux must  have the strict ``Freund-Rubin'' form: 
\begin{equation}
F^{(d)} ~=~ f \, {\rm Vol}_{{\rm AdS}_d} ~+~ F^{(d)}_{\rm internal} \,,
\end{equation}
where $f$ is a {constant} and $F^{(d)}_{\rm internal}$ only has components along the internal manifold.  Thus, if one considers M2 or D3 branes parallel to the boundary of the Poincar\'e patch of AdS, the electric flux potential that couples to the branes is universal and independent of their location in the internal manifold.

Hence, the action of a probe brane will generically depend on its position on the internal manifold, and this dependence {\it only} comes from the DBI part of the action.  As we will show, this can create regions, ${\cal J}$,  in which the gravitational attraction is less than the electrostatic repulsion:  inside this ``jet locus,''  ${\cal J}$, the brane probe will be ejected from the AdS.

It is perhaps tempting to try to interpret the brane probe analysis of \cite{Maldacena:1998uz, Ooguri:2016pdq} as referring to smeared M or D branes.  However, fundamental brane probes always have equal charge and tension and so smearing will not change this. One might try to interpret the warp factor contribution to the DBI action as creating some kind of  effective tension but, as we noted above, if one averages the actions of these branes over the internal manifold, the brane-jet instabilities disappear and  the smeared potential is always attractive. 

There are two other intuitive ways in which one may try to link the brane-jet instability to the WGC and to the arguments that lead to the AdS instability conjecture of \cite{Ooguri:2016pdq}. One is to argue that all the non-supersymmetric AdS$_4$ solutions of gauged supergravities that come from M theory are in fact the result of a geometric transition of a system of M2 branes on top of a 2+1 dimensional object of M theory that breaks supersymmetry\footnote{This is only understood for a few examples and is far from obvious in general.}. As such, by the WGC, some of the M2 branes should acquire charge larger than the mass (in the units of mass and charge of the background) and should run away from the other branes\footnote{We thank C. Vafa for discussions leading to this interpretation}, giving rise to the brane-jet instability. One piece of evidence in support of this intuitive picture is that the radial dependence of the repulsive force is the same as the radial dependence of the repulsive force between M2-branes whose charge is larger than the mass. However, despite this non-trivial agreement, we find that several non-supersymmetric AdS vacua do not have brane-jet instabilities (see,  Table 1 of Section 4), but rather decay through modes whose mass is below the BF bound.

The second way is to imagine compactifying the AdS$_4$ solution along the two space-like boundary directions of the Poincar\'e patch and obtain an effective 1+1 dimensional theory. The particles in this theory come from M2 branes extended along the compactified AdS directions and, for non-trivial $\Delta$, their masses will depend on their location on the internal seven-manifold. If one thinks about M2 branes at different locations as giving rise to different species of particles in the 1+1 dimensional theory, one expects by the WGC that some of these particles will have a mass smaller than the charge and will thus be expelled away from the Poincar\'e horizon. These particles would thus descend from the M2 branes localized in the region of the jet. Much like the previous intuitive picture, the dependence of the force on the distance is consistent to what one expects from the WGC, but the absence of this instability in other non-supersymmetric AdS vacua indicates that this intuitive picture should be taken with a grain of salt.

Ultimately it seems rather unnatural to force the brane-jet instability into the framework of \cite{Maldacena:1998uz, Ooguri:2016pdq}. The underlying physics in this paper is relatively simple.  First, we  are using fundamental brane probes that, by definition, have tension equal to charge.  The AdS$_4$ solutions we explore can be thought of as coming from the back-reaction of M2 branes that have non-trivial magnetic fields on their ``internal dimensions,'' which means magnetic fields that thread the sphere surrounding the branes in eleven dimensions. We are showing that, from the eleven-dimensional perspective, the strong magnetic flux deforms the gravitational field on the sphere around the M2 branes  in such a manner as to weaken the gravitational field in some regions and strengthen  it in others. As a result, the stack of magnetized M2 branes attract fundamental brane probes in some directions and squirt them out in others.  While the physical mechanism is rather different,  the behaviour of a brane jet has significant commonality with its astrophysical cousins.

It is interesting to understand why the brane-jet instabilities dodge the standard supergravity stability theorems, and how one might find these instabilities within the traditional Kaluza-Klein excitations of the supergravity theory.   Supersymmetry generally gives some protection against instabilities through positive mass theorems and BPS bounds.  On the other hand, 
the general problem with analyzing stability for non-supersymmetric solutions is that there are typically some very significant assumptions about the decay channels.  For the stability analysis of gauged supergravity, most of the theorems involve  perturbations that lie within a restricted set of supermultplets\footnote{This is why it can be misleading to refer to the  $SO(3) \times SO(3)$ invariant, AdS$_4$ vacuum as  BF-stable:  BF stability has only been verified for the scalars in the maximal graviton multiplet.}.  Similarly, the positive mass theorems are also usually restricted to the fields in the massless graviton supermultiplet.   The problem is that the fields in the massless graviton supermultiplet of the maximally supersymmetric theory in  lower dimensions, correspond to the ``lowest harmonics'' on the sphere, which necessarily vary very slowly over the sphere. In contrast, the brane jet instability can involve the emission of branes that are much more localized on the internal directions.   

There should, however, be an extensive overlap between what one can see using  brane probes and what one can see using supergravity scalars. When branes spread out onto a Coulomb branch, this involves giving expectation values to the scalar fields from the perspective of the lower-dimensional supergravity.  Indeed, there are plenty of examples of holographic renormalization group flows that describe the spreading of branes out onto a Coulomb branch\footnote{See, for example, \cite{Freedman:1999gk,Carlisle:2003nd,Gowdigere:2005wq, Skenderis:2006di}.}.    With the exception of supersymmetric vacua, the stability theorems involving scalars in supergravity are typically restricted to the massless graviton supermultiplet, and hence to the lowest harmonics on the compactification manifold, or collective effects of these lowest modes as captured by gauged supergravity.  Such theorems are ill-adapted to studying the emission of individual branes that localize on the compactification manifold.  At the very least, such branes would excite higher order Fourier modes. Moreover, some brane instabilities might require the collective effect of many such modes to replicate the brane dynamics. 

Thus  brane-jet instabilities and supergravity scalar perturbations are really probing similar instabilities but from very different perspectives.  The bottom line is that, if there is a localized brane-jet instability, then one must look at higher order scalar harmonics: the Fourier expansions must be able to resolve ${\cal J}$ if it is to detect the repulsion.   Moreover, some brane-jet instabilities may require non-trivial combinations of modes so that the repulsion from ${\cal J}$ is not washed out by the attraction from  ${\cal J}^c$. 
 
There is now a wealth of literature that explores other possible instabilities, especially through instantons (see, for example \cite{Witten:1981gj,Maldacena:1998uz,Horowitz:2007pr,Narayan:2010em,Ooguri:2017njy,Apruzzi:2019ecr}  and the references therein). There are also some interesting results on the construction of non-supersymmetric AdS$_4$ vacua \cite{Borghese:2012qm,Guarino:2015vca} in dyonically-gauged maximal supergravity \cite{DallAgata:2011aa,DallAgata:2014tph, Guarino:2015qaa}, and the stability of AdS vacua in massive IIA supergravity \cite{Gaiotto:2009mv,Danielsson:2017max}. 

In Section 2 we briefly review our conventions and some of the relevant details of the relationship between the scalar fields and potentials in four dimensions and the Freund-Rubin flux and warp factors in eleven dimensions.  In Section 3 we introduce the brane probe analysis and study brane jets analytically for the most symmetric AdS$_4$ vacua that were found in \cite{Warner:1983du,Warner:1983vz}.  In particular, we show that the $SO(3) \times SO(3)$ invariant vacuum of gauged $\cN =8$ supergravity in four dimensions, which is BF stable in the gauged $\cN =8$ theory,  has a  brane-jet instability.  We will show that, for this solution, the  jet locus, ${\cal J}$, is very small and has a relatively weak repulsive force when compared to the typical attractive force on brane probes that localize outside of  ${\cal J}$.  This means that if one smears the M2 branes over the internal manifold, the instability disappears.  Thus the brane localization in a ``jet'' is critical to this instability.

In Section 4 we catalogue the brane-jet instabilities of a sample of 26 AdS$_4$ vacua, most of which are non-supersymmetric and BF unstable.  We show that many (but not all) of them have brane-jet instabilities and we show that the repulsion in the brane-jet region is also typically weak and that the averaged (smeared) probe potential is attractive. 

Our final comments are in Section 5 and the appendices contain further relevant details of the uplift formulae that are essential to reconstructing the eleven-dimensional metric from the four-dimensional data.

\section{Conventions and normalizations} 
\label{sec:Conventions}

Since the dynamics of M2 brane probes is going to depend critically on the balance of electrostatic and gravitational forces, it is going to be crucial to start by getting our conventions and normalizations fixed.

\subsection{Eleven dimensions}
\label{ss:11D}

First, we are going to work in old supergravity conventions for M theory.  The equations of motion are thus:
\begin{equation}
\begin{aligned}
R_{M N} ~+~ R \, g_{M N}  ~=~&  \coeff{1}{3}\,   F_{MPQR }\, F_N{}^{PQR}\,, \\[6 pt]  
\nabla_M F^{MNPQ} ~=~& -  \coeff{1}{576} \, \varepsilon^{NPQ  R_1 R_2 R_3  R_4 S_1S_2  S_3  S_4} \, F_{ R_1 R_2 R_3  R_4} \, F_{ S_1S_2  S_3  S_4} \,, 
 \end{aligned}
 \label{eoms}
 \end{equation}
with  $\tilde \varepsilon^{1\cdots 11} = 1$.  The supersymmetry transformation of the gravitino is:
\begin{equation}
\delta \psi_M ~=~ \nabla_M \, \epsilon ~+~ \coeff{1}{144}\,
\Big({\Gamma_M}^{N PQR} ~-~ 8\, \delta_M^N  \, 
\Gamma^{PQR} \Big)\, F_{NPQR} \, \epsilon~=~ 0 \,.
 \end{equation}
The eleven-dimensional metric will be ``mostly plus'' and taken to have the form
\begin{equation}
ds_{11}^2 ~=~ \Delta^{-1} \, \widehat{ds}{}^2_{1,3}~+~  ds_7^2 \,.
 \end{equation}
The coordinates on the four-dimensional space-time will be denoted by $x^\mu$ while the coordinates on the internal manifold will be denoted by $y^m$. This internal manifold will always be topologically $S^7$ but its metric:
\begin{equation}
ds_{7}^2 ~=~  g_{mn} dy^m \, dy^n \,,
 \end{equation}
will generically be deformed away from the round metric.

The AdS$_4$ metric, $\hat g_{\hat \mu \hat \nu}$, will be written in the  Poincar\'e patch:
\begin{equation}
\widehat{ds}{}^2_{1,3} ~=~ \hat g_{\hat \mu \hat \nu}\, dx^{\hat \mu}  dx^{\hat \nu} ~=~ e^{2 A(r)} \eta_{\mu\nu}\,  dx^\mu dx^\nu ~+~ dr^2 \,,
 \end{equation}
where $x^{\hat \mu} = (x^\mu, r)$ and 
\begin{equation}
A(r) ~=~r/L_* \,,
 \end{equation}
which defines the radius, $L_*$, of the metric $\hat g_{\hat \mu \hat \nu}$.

The flux, $F$, will be decomposed into is components along the AdS and the internal manifold according to:
\begin{equation}
F ~=~ F^{({\rm st})} ~+~ F^{({\rm int})}  \,,
 \end{equation}
and we will define the Freund Rubin constant, $\mathfrak f_\text{FR}$, by setting
\begin{equation}
 F^{({\rm st})} ~=~  \mathfrak f_\text{FR} \, e^{3 A(r)} \, dx^0  \wedge dx^1  \wedge dx^2 \wedge dr \,.
 \label{AdSvol}
 \end{equation}
Note that this means that the electric components of the $3$-form potential are given by: 
\begin{equation}
 A^{({\rm st})} ~=~  - \frac{\mathfrak f_\text{FR}\,L_* }{3} \, e^{3 A(r)} \, dx^0  \wedge dx^1  \wedge dx^2 \,.
 \label{Ast-res}
 \end{equation}

There is a potential source of confusion here when the warp factor, $\Delta$, is a constant.  Note that the spatial components of the full metric are $g_{ \mu  \nu}  = \Delta^{-1} \hat g_{\hat \mu \hat \nu} $, which is also AdS$_4$ for constant $\Delta$.  It is important to remember that the AdS$_4$ volume form in (\ref{AdSvol}) is that of the four-dimensional metric, $\hat g_{\hat \mu \hat \nu}$, and not the volume form of $g_{ \mu  \nu}$.  If one is not careful about this, then one is apt to get spurious powers of $\Delta$ in $\mathfrak f_\text{FR}$.

The round metric on $S^7$ will be denoted by $\go_{mn}$, and the inverse  radius of this round sphere will always be denoted by $m_7$.   
The round Ricci tensor is thus:
\begin{equation}
 \Ro_{mn} ~=~ 6\, m_7^2 \,  \go_{mn}\,.
 \end{equation}

The determinant factor, $\Delta$, will always be normalized against the round sphere solution:
\begin{equation}
\Delta ~\equiv~ \sqrt{ \frac{\det(g_{mn})}{\det(\go_{mn})}} \,.
\label{Deltadefn}
 \end{equation}

For the solution based on the round $S^7$, the corresponding AdS factor has 
\begin{equation}
 \Ro_{\hat\mu \hat\nu} ~=~ -3\, m_4^2 \,  \go_{\hat\mu \hat\nu}\,,
 \label{roundRic4}
 \end{equation}
where $m_4$ is the inverse radius of the AdS$_4$.  The parameters of this solution are related by: 
\begin{equation}
m_4 ~=~ 2\, m_7 \,, \qquad  \mathfrak f_\text{FR} ~=~ - \coeff{3}{2}\, m_4 ~=~   -3 \, m_7 \,.
\label{roundparams}
 \end{equation}

In our discussion, the scale, $m_7$, will always be that of the round sphere.  The parameters $L_* = m_4^{-1}$ and $\mathfrak f_\text{FR}$ will vary from solution to solution.

\subsection{Four dimensions}
\label{ss:4D}

The dynamics in four dimensions is governed by the scalar potential, $\cP$.  The  AdS$_4$ vacua of interest correspond to critical points of $\cP$ and the Einstein equations reduce to: 
\begin{equation}
 \hat R_{\hat\mu \hat\nu} ~=~   \cP \,  \hat g_{\hat\mu \hat\nu}~=~ -\frac{3}{L_*^2} \,  \hat g_{\hat\mu \hat\nu}
 \end{equation}
The potential is proportional to $g^2$, where $g$ is the gauge coupling constant of the gauged supergravity theory. Since $\cP$ is negative at all known critical points, it is convenient to define 
\begin{equation}
P_* ~\equiv~  -\frac{\cP}{g^2}  \,.
 \label{Pstar}
 \end{equation}
At the maximally symmetric critical point, which corresponds to the round $S^7$, one has $\cP = -6 g^2$, and comparing this to (\ref{roundRic4}) and (\ref{roundparams}) yields the universal relationship:
\begin{equation}
g ~=~  \sqrt{2} \, m_7 \,.
 \label{gm7reln}
 \end{equation}
and so, at an arbitrary critical point one has 
\begin{equation}
L_*  ~=~  \sqrt{\frac{3}{2\, P_*}} \, m_7^{-1} \,.
 \label{Lreln}
 \end{equation}
%

\subsection{Uplift results}
\label{ss:Uplift}

There is a proven relationship between the scalar matrices in four dimensions and the inverse metric in eleven dimensions \cite{deWit:1984nz}. We are not going to go into this here.  Indeed, all we will need is the expression for $\Delta$ defined in (\ref{Deltadefn}) and we will simply take the results from the existing literature when we need them.  Further details can be found in Appendix \ref{App:Uplifts}.

We will, however note that there is an well-established empirical relationship \cite{Nicolai:2011cy,Godazgar:2013nma} between the Freund-Rubin parameter, as defined above, and the critical value of $\cP$.  That is, one has:
\begin{equation}
\mathfrak f_\text{FR} ~=~ -\frac{1}{2} \, m_7\,  P_* \,.
 \label{fFRreln}
 \end{equation}
We will need the coefficient of the gauge  potential in (\ref{Ast-res}) and this is given by:
\begin{equation}
- \frac{\mathfrak f_\text{FR}\,L_* }{3}  ~=~ \frac{1}{2} \, \sqrt{\frac{P_*}{6}}\,.
 \label{Acoeff}
 \end{equation}
This means that the strength of the electric repulsion of M2 branes is universally determined by the value of the potential at its critical point.

\section{M2-brane probes and the most symmetric AdS vacua} 
\label{sec:probes}

The M2-brane probe action in our conventions is:
\begin{equation}
S ~=~  \int \, \Big[  \sqrt{- \det(\tilde g)} \, d^3 \sigma ~+~ 2\, \coeff{1}{3!} \, \tilde A^{(3)}\Big] \,,
 \end{equation}
where $\tilde g$ and $\tilde A^{(3)}$ denote the pull-back of the metric and the $3$-form onto the membrane.  
Since we are using old supergravity conventions that lead to equations of motion in the form (\ref{eoms}),   the normalization of the $A^{(3)}$-term here is twice the more common ``stringy'' convention.    

We consider a probe that is parallel to the boundary of the Poincar\'e patch and we are only interested in the potential energy of the brane, and so it suffices to take the brane embedding to be $x^\mu = \xi^\mu$.  The resulting brane potential is therefore
\begin{equation}
V ~=~  e^{3 A(r)} \, \bigg (\,\Delta^{-\frac{3}{2}}  ~-~ \sqrt{\frac{P_*}{6}} \, \bigg) \,,
 \end{equation}
where we have used (\ref{Acoeff}) to replace $\mathfrak f_\text{FR}$.  Note that because $A(r) = r/L_*$, one has 
\begin{equation}
\frac{ d V}{dr}  ~=~  \frac{3}{L_*} \,e^{3 A(r)} \, \bigg (\,\Delta^{-\frac{3}{2}}  ~-~ \sqrt{\frac{P_*}{6}} \, \bigg) \,,
 \end{equation}
which is positive or negative depending on the sign of 
\begin{equation}
\Theta ~\equiv~ \bigg (\,\Delta^{-\frac{3}{2}}  ~-~ \sqrt{\frac{P_*}{6}} \, \bigg) \,.
 \end{equation}
This is exactly as it should be because, if $\Delta^{-\frac{3}{2}} > \sqrt{\frac{P_*}{6}}$, the gravitational attraction felt by the brane is larger than electrostatic repulsion, and the force must be attractive toward $r=0$ ($\frac{d V}{dr} >0$).  Conversely,    if $\Delta^{-\frac{3}{2}} <  \sqrt{\frac{P_*}{6}}$ then the electrostatic repulsion wins over gravity, and the force must be repulsive  ($\frac{d V}{dr} < 0$).

In the following sections  we are going to calculate the ``average,'' $\Theta_\text{avg}$, of $\Theta$ over $S^7$.  This average will always be taken by integrating over the round $S^7$ metric, with a measure factor of $\sqrt{\det(\go_{mn})}$.   

Note that at the maximally symmetric critical point one has $\Delta \equiv 1$ and $P_* = 6$.  The potential thus vanishes identically, indicating that every direction is ``flat'' and the brane probe feels no force anywhere.  This is, of course, because the probe is mutually BPS with the stack of M2 branes whose near-horizon geometry the maximally supersymmetric AdS$_4 \times S^7$ solution is.

We now consider non-trivial critical points, which correspond to turning on background magnetic fields in eleven dimensions.  In this section we will examine the most symmetric AdS$_4$ vacua that were first given in \cite{Warner:1983du,Warner:1983vz}. While we will provide the original references for the relevant data, the work of Kr\"uger  is also an excellent  resource \cite{Kruger:2016agp,Kruger:2016yin}.   We will start with the non-trivial supersymmetric point.

\subsection{The $SU(3) \times U(1)$ invariant point}
\label{ss:SU3}

This critical point has been extensively studied and it represents a conformal fixed point at the end of a non-trivial holographic holographic RG flow from the ABJM theory.  The properties of this critical point have been extensively analyzed in four-dimensional supergravity in  \cite{Nicolai:1985hs}. The holographic flow solutions were first constructed in four dimensions in \cite{Ahn:2000aq,Ahn:2000mf}.  This flow was uplifted to M theory in \cite{Corrado:2001nv}. The  holographic interpretation in ABJM theory is given in \cite{Klebanov:2008vq}.  The important point here is that this fixed point corresponds to integrating out one massive chiral multiplet, leaving three massless multiplets at the fixed point.  These give rise to a six-dimensional (three complex dimensional) Coulomb branch at the fixed point.

The brane probe calculation has already been done for the entire flow in \cite{Corrado:2001nv} and so we simply repeat the results here, and only for the fixed point.
The result is:
\begin{equation}
\Theta ~= 3^{\frac{3}{4}} \, \sin^2 \mu  ~\ge ~ 0  \,,
\label{SU3pot}
 \end{equation}
where $\mu$ is the coordinate that defines how the $\IC\IP^2$, on which the $SU(3)$ is transitive, sits in the round $S^7$.  Specifically, the metric on the round $S^7$ may be written:
\begin{equation}
ds_7^2 ~=~ ( d\psi+\cos^2\mu\,(d\phi+A))^2~+~  d\mu^2 ~+~ \cos^2\mu \,\big(d_{FS(2)}^2 + \sin^2\mu\, (d\phi +A)^2\big)  \,,
\label{SU3met}
 \end{equation}
where $d_{FS(2)}^2$ is the usual Fubini--Study metric on $\IC\IP^2$ and $A$ is the $U(1)$ connection of the Hopf fiber over $\IC\IP^2$.

The potential, (\ref{SU3pot}), is attractive except when $\mu=0$.   From (\ref{SU3met}), one sees that vanishing locus  of the potential are five-sphere sections of $S^7$, which, when combined with the radial coordinate, define the Coulomb branch of the infra-red fixed point.  The metric on this moduli space is, of course, induced from that of the deformed $S^7$ of the compactification and was computed in \cite{Corrado:2001nv}.

\subsection{The  $G_2$, $SO(7)^+$ and $SO(7)^-$  invariant points}
\label{ss:G2SO7}

The $SO(7)^\pm$ points are both non-supersymmetric and are known to be BF unstable while the $G_2$ point has $\Neql1$ supersymmetry.  

Define the coordinate $\theta$ by writing the metric of the round $S^7$ in terms of the $S^6$ upon which the $SO(7)^+$ acts transitively:  
\begin{equation}
ds_7^2 ~=~  d \theta^2 ~+~ \sin^2 \theta \, d \Omega_6^2  \,. 
\label{thetadefn}
\end{equation}
Then, using the results of \cite{deWit:1984nz}, we find:
\begin{equation}
\begin{aligned}
\Theta_{G_2}~=~ & \frac{2^{\frac{7}{4}} 3^{\frac{13}{8}} }{5^{\frac{7}{4}} }\,\bigg[3 - \sqrt{\frac{5}{6}} ~-~ 2 \sin^2 \theta \bigg] ~>~ 0\,, \\
\Theta_{SO(7)^+}~=~  &5^{-\frac{1}{8}} \, \bigg[5 - \sqrt{\frac{5}{3}} ~-~ 4 \sin^2 \theta \bigg] \,, \\
\Theta_{SO(7)^-}~=~  &\frac{5^{\frac{5}{2}} }{4 \sqrt{3}}\, \bigg[\frac{1}{2} \sqrt{\frac{15}{2}} ~-~ 1  \bigg] ~>~0\,.
\end{aligned}
\label{}
\end{equation}
%

\begin{figure}
\leftline{\hskip 3 cm \includegraphics[width=9cm]{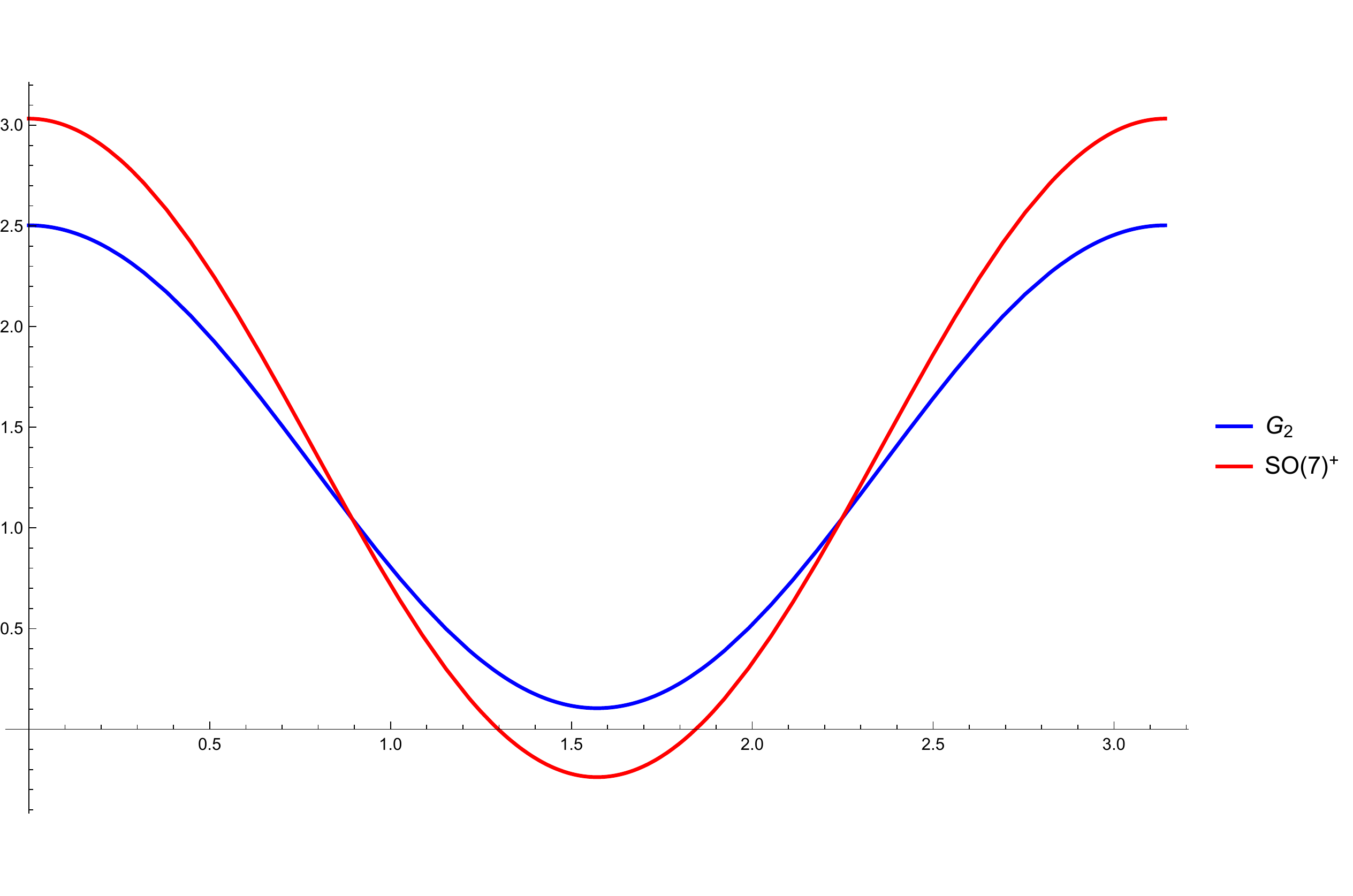}}
\setlength{\unitlength}{0.1\columnwidth}
\caption{\it 
Plot of $\Theta (\theta)$ for the $G_2$ (in blue) and $SO(7)^+$ (in red)  critical points.  Note that for $G_2$, $\Theta$ remains strictly positive while for $SO(7)^+$,   $\Theta$ becomes negative in a region around $\theta = \frac{\pi}{2}$. This negative region is significantly larger than it appears in this plot because of the volume measure factor in (\ref{normvol}).}
\label{fig:G2SO7-plots}
\end{figure}

The action of $SO(7)^-$ is transitive on $S^7$ and so there is no dependence on position. The coefficient $\Theta$ is therefore simply a constant, and is positive.  The M2 brane probes are thus attracted uniformly in this solution.  Indeed, as we will see, if one averages over the $S^7$, we will see that brane probes are generically attracted.  This solution has too high a level of symmetry to see anything but the ``averaged'' attractive force.

The other two solutions have non-trivial $\Theta$.  The potential for the $G_2$ point is always attractive and this is consistent with the stability one expects of supersymmetric solutions.  This potential has  no flat directions, which means that there isn't a Coulomb branch upon which the branes can spread with ``zero force.''

The $SO(7)^+$ point exhibits a brane-jet instability in a region, $\cJ$, around $\theta = \frac{\pi}{2}$. (See Fig.~\ref{fig:G2SO7-plots}.) Because the volume measure is hugely peaked near $\theta = \frac{\pi}{2}$, the plots in Fig.~\ref{fig:G2SO7-plots} give a somewhat distorted view of the region in which  $\Theta_{SO(7)^+}$ is negative.  To compute the ``area'' of $\cJ$,  note that the  measure on the round $S^7$ described by (\ref{thetadefn}) is $\sin^6  \theta$ with $0\le \theta \le \pi$.  The normalized volume integral and the relative ``area'' of $\cJ$ are therefore given by:
\begin{equation}
\frac{16}{5\pi} \, \int_0^{\pi}  \, \sin^6  \theta\,    d \theta   ~=~ 1 \,, \qquad  \frac{16}{5\pi} \, \int_{\cJ}  \, \sin^6  \theta\,    d \theta   ~\approx~ 0.51723818 \dots  \,.
\label{normvol}
\end{equation}
Thus the jet domain  actually  covers slightly more than half of the $S^7$.

The  average of $\Theta$ for the   $SO(7)^+$ point is given by 
\begin{equation}
\frac{16}{5\pi} \, \int_0^{\pi}  \, \Theta(\theta) \, \sin^6  \theta\,    d \theta   ~=~ \frac{3}{2}  \, 5^{-\frac{1}{8}}\,\bigg[\,1  - \frac{2}{3} \sqrt{\frac{5}{3}}\,\,  \bigg]  ~\approx~   0.1709175 \dots  \,.
\label{SO7avg}
\end{equation}
and so the average force is still weakly attractive. 

One could also choose to average over the metric of the deformed $S^7$ at the critical point.  From (\ref{Deltadefn}), one sees that this would involve inserting an extra factor of $\Delta$ into the  round $S^7$ metric measure.   We then find that the area of the deformed $S^7$ is $0.930334963898 .. $ times the round value, while the integral (\ref{SO7avg}) with an extra factor of $\Delta$  is $0.186473804279 ... $, which means that the average of $\Theta$ over the deformed $S^7$ is $0.200437274224 ... $, which is not very different from the average over $S^7$.

\subsection{The $SU(4)^-$  invariant point}
\label{ss:SU4}

The values of $\Delta$ and  $\mathfrak f_\text{FR}$ can be obtained from \cite{Pope:1984bd} after carefully rescaling the eleven-dimensional solution according to the conventions described in Section  \ref{sec:Conventions}.

Like $SO(7)^-$, the action of $SU(4)^-$ is also transitive on $S^7$ and so there is no dependence on position. The coefficient $\Theta$ is therefore simply a constant, and is positive:
\begin{equation}
\Theta_{SU(4)^-}~=~ 2 ~-~ \frac{2}{\sqrt{3}}  ~>~0\,.
\label{}
\end{equation}
Again, because of the high level of symmetry, one is only seeing the ``averaged''  force, and this is attractive.

\subsection{The $SO(3) \times SO(3)$  invariant point}
\label{ss:SO3SO3}

The complete uplift of the $SO(3)\times SO(3)$-invariant critical point to M theory was obtained in  \cite{Godazgar:2014eza}.  The metric here is considerably more involved than for the critical points discussed above because the orbits of the $SO(3)\times SO(3)$ isometry of the internal metric  have co-dimension two (rather than zero or one) on the deformed $S^7$,  and are  isomorphic with the coset
\begin{equation}\label{}
T^{1,1}~\equiv~{SU(2)\times SU(2)\over U(1)}\,. 
\end{equation}
A natural set of the internal coordinates, $y^m$, consists  of five Euler angles   and two transverse coordinates, $\rho$ and $\varphi$, where
$0\leq \rho\leq {\pi\over 2}$ and $ 0\leq\varphi\leq 2\pi$.
The  invariant forms, $\sigma_{(i)}^{1,2,3}$, $j=1,2$, for each $SU(2)$, become linearly dependent when pulled back onto  $S^7$, such that 
\begin{equation}\label{so3onef}
\sigma^1_{(1)}\,,\quad \sigma^2_{(1)}\,,\quad \sigma^1_{(2)}\,,\quad \sigma^2_{(2)}\,,\quad \sigma^3\equiv\sigma^3_{(1)}-\sigma^3_{(2)}\,,
\end{equation}
yield a local frame along the orbit. In this formulation,  the round metric on $S^7$ is given by
 \begin{equation}\label{}
d\so ^2_7 = {\frac{1}{4m_7^2}}\,\Big[ d\rho^2+\sin^2\rho\,d\varphi^2 +\big(\sigma_{(1)}^+\sigma_{(1)}^-+\sigma_{(2)}^+\sigma_{(2)}^-\big)
-\sin\rho\,\big(\sigma_{(1)}^+\sigma_{(2)}^-+\sigma_{(2)}^+\sigma_{(1)}^-\big)+\big(\cos\rho\,d\varphi-\sigma^3\big)^2\,
\Big]\,,
\end{equation}
where $\sigma_{(j)}^\pm= \sigma_{(j)}^1\pm i\sigma^2_{(j)}$.
The actual internal metric for the uplift  has the same invariant one-forms \eqref{so3onef}, but more complicated coefficient functions of $\rho$ and $\varphi$. 
The  warp factor can be written as
\begin{equation}\label{warpsimp}
\Delta^{-3}\eql {1\over 36}\,\Big[ \left(2\xi-3\sqrt 5\right)^2+10\,\left(2\xi-3\sqrt 5\right)\left(2\zeta-3\sqrt 5\right)+ \left(2\zeta-3\sqrt 5\right)^2\Big]\,,
\end{equation}
where
$\xi\equiv 3\,\sin\rho\cos\varphi$ and $ \zeta\equiv 3\,\sin\rho\sin\varphi$ are two $SO(3)\times SO(3)$ invariant functions on $S^7$. 

A straightforward calculation using $P_*=14$ gives  
\begin{equation}
\Theta(\rho,\varphi) =   \sqrt{\frac{31}{2}-\frac{1}{2}\cos (2 \rho ) - 6\,\sqrt{10} \, \cos  (\varphi + \alpha)\,\sin  \rho + 5  \cos (2  (\varphi + \alpha) ) \,\sin^2  \rho } -\sqrt{7\over 3}
\,.
\label{}
\end{equation}
where we have restored an arbitrary parameter $\alpha$ that parametrizes a family of inequivalent  $SO(3)\times SO(3)$ invariant  critical points \cite{Bobev:2011rv}\footnote{The parameter $\alpha$ corresponds to an $U(1)$ symmetry of the scalar  potential that lies outside the $SO(8)$ gauge group. The warp factor for a general $\alpha$ is given in \cite{Godazgar:2014eza} Eqs.\ (5.12). A convenient ``symmetric'' choice is $\alpha=-\pi/4$, which yields \eqref{warpsimp}. }. 
This has a minimum value of $\sqrt{2} - \sqrt{\frac{7}{3}} \approx -0.1133$ at $(\rho,\varphi)= (\frac{\pi}{2}, -\alpha \pm \arctan(\frac{1}{3}))$  and a maximum value of $\sqrt{6}+\sqrt{15}  - \sqrt{\frac{7}{3}} \approx  4.7949$ at $(\rho,\varphi)= (\frac{\pi}{2}, -\alpha + \pi)$.

\begin{figure}
\leftline{ \hskip 3 cm\includegraphics[width=9cm]{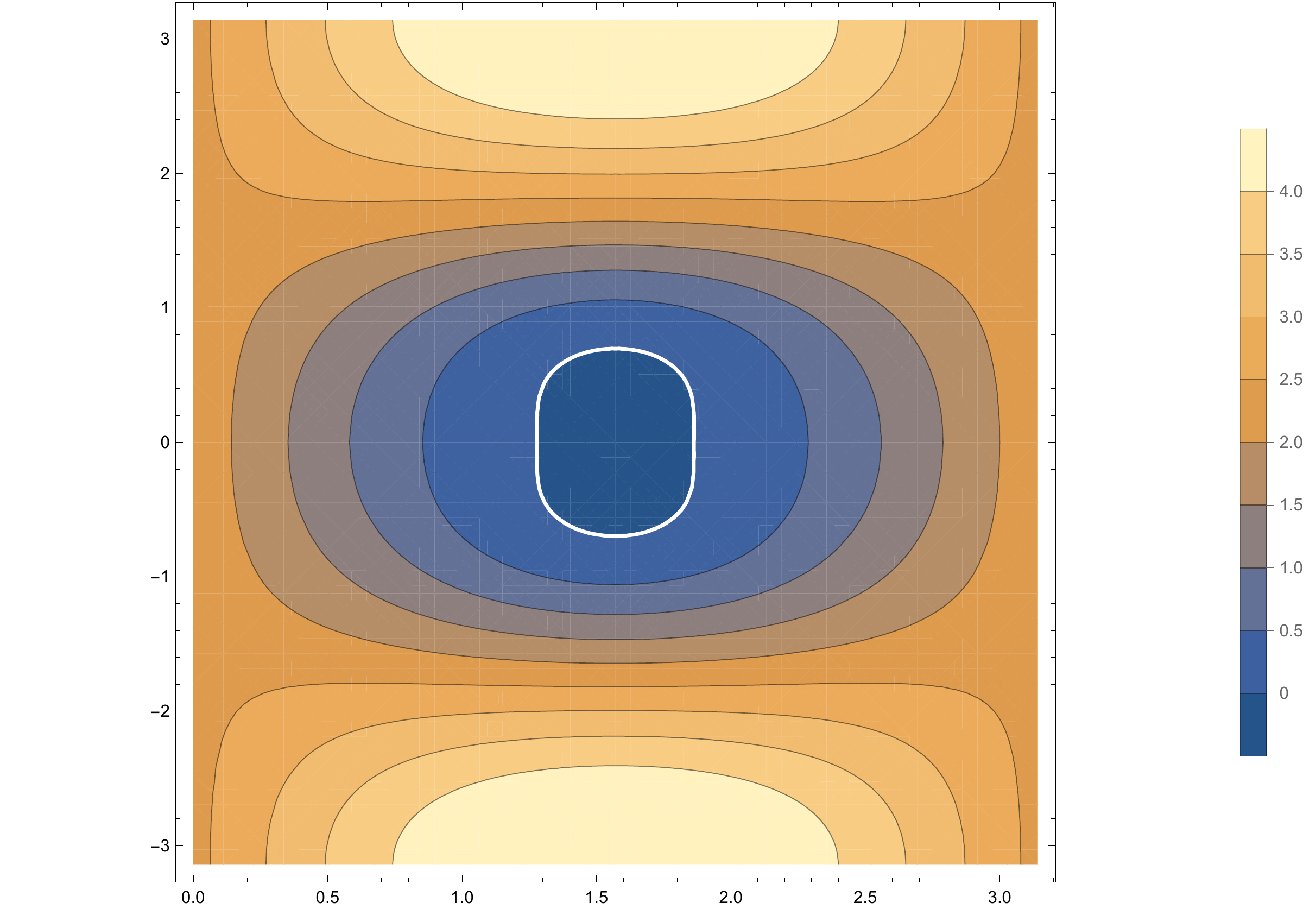} }
\setlength{\unitlength}{0.1\columnwidth}
\caption{\it 
Contour plot of $\Theta (\rho,\varphi)$ with $\alpha =0$ for the $SO(3) \times SO(3)$ critical point. The brane-jet locus, $\cJ$, is the interior of the region defined by the bold white line.  Note that while this region appears to have  a moderate size, this is a distortion caused by the projection. The region, $\cJ$ actually occupies slightly less than $0.4\%$ of the $S^7$.   The minimum value of $\Theta$ is $\sqrt{2} - \sqrt{\frac{7}{3}} \approx -0.1133$,  and the maximum value of $\Theta$ is $\sqrt{6}+\sqrt{15}  - \sqrt{\frac{7}{3}} \approx 4.7949$.  The average of $\Theta$ over $S^7$ is approximately $2.230758$. Thus the brane-jet instability is very shallow and very localized. }
\label{fig:SO3SO3contours}
\end{figure}

The contours of this potential are shown in Fig. \ref{fig:SO3SO3contours}, in which the jet domain, $\cJ$, is clearly delineated.  One should note that the actual size of $\cJ$ is much smaller than it appears.  This is the usual map projection issue: the measure  (\ref{volform1}) vanishes quadratically as $\rho \to \frac{\pi}{2}$ and so the true area of $\cJ$ is greatly compressed.  Numerical integration reveals that $\cJ$ actually covers less than $0.4\%$ of the round $S^7$ area\footnote{This is about the same as the area of Greenland as a fraction of the Earth's surface.}. 

At fixed values of the coordinates $\rho$ and $\varphi$ there is an $S^2$  and an $S^3$, non-trivially fibered.  The measure is:
\begin{equation}
\coeff{1}{16} \,(4 \pi)\, (2 \pi^2) \cos^2 \rho\,  \sin \rho \, d\rho \, d\varphi ~=~\coeff{1}{2} \,\pi^3\, \cos^2  \rho  \sin  \rho\, d\rho \, d\varphi  \,,
\label{volform1}
\end{equation}
where the factors of $4 \pi$ and $2 \pi^2$ are the volumes of the unit  $S^2$ and an $S^3$.  The integration range is $\rho \in [0, \frac{\pi}{2}]$ and $\varphi \in [0, 2\pi]$.   Note that the volume of $S^7$ is $ \frac{1}{3} \pi^4$ and thus the average of $\Theta(\rho,\varphi)$ over $S^7$ is done by taking:
\begin{equation}
\frac{3}{2\pi} \, \int_0^{ \frac{\pi}{2}} d \rho\, \int_0^{ 2\pi} d \varphi \, cos^2  \rho  \sin  \rho \, \Theta(\rho,\varphi)  \,,
\label{intavg}
\end{equation}
Numerical integration gives this average value as $2.2307583736$.  So the average force is strongly attractive, and twenty times larger than the minimum value.

As above, one could also choose to average over the metric of the deformed $S^7$ at the critical point.   We then find that the area of the deformed $S^7$ is $0.4381475058\dots$ times the round value, while the integral (\ref{intavg}) with an extra factor of $\Delta$  is $0.868758220 \dots $, which means that the average of $\Theta$ over the deformed $S^7$ is $1.982798507 \dots$, which is, again, not significantly different from the average over the round $S^7$.  We will therefore continue the practice of  averaging $\Theta$ using the round metric measure. 

\section{More AdS vacua:  summary and numerical estimates} 
\label{sec:NEstimates}

\begin{table}[t]
\renewcommand{\arraystretch}{1.0}
\begin{center}
\resizebox{0.95\textwidth}{!}{%
\begin{tabular}{@{\extracolsep{8 pt}}  c c c c c c c c c c c c}
\toprule
\noalign{\smallskip}
Point & Symmetry & Susy  & $\Delta^{-3}_\text{max}$ & $\Delta^{-3}_\text{min}$ & $P_*/6$ & $\Theta_\text{min}$ & $\Theta_\text{avg}^{\le}$ & BF & BJ \\
\noalign{\smallskip}
\midrule
\noalign{\smallskip}
{\tt S0600000} & SO(8) & $\mathcal{N}=8$ & 1 & 1 & 1 & 0 & 0 & S & S
\\
{\tt S0668740} & SO(7)$^+$ & -- & 16.718 & 0.6687 & 1.1146 & $-0.2380$ & 0.1213 & U & U
\\
{\tt S0698771} & SO(7)$^-$ & -- & 2.1837 & 2.1837 & 1.1646 & +0.3986 & 0.3986 & U & S
\\
{\tt S0719157} & G$_2$ & $\mathcal{N}=1$ & 12.945 &  1.4383 & 1.1986 & +0.1045 & 0.2503 & S & S
\\
{\tt S0779422} & SU(3)$\times$U(1) & $\mathcal{N}=2 $ & 11.691 &  1.2990 & 1.2990 & 0 & 0.3989 & S & S
\\
{\tt S0800000} & SU(4)$^-$ & -- & 4.0000 & 4.0000 & 1.3333 & $+0.8453$ & 0.8453 & U & S
\\
{\tt S0869596} & SO(3)$\times$U(1) & -- & 27.230 & 0.7427 & 1.4493 & $-0.3421$  & 0.4869 & U & U
\\
{\tt S0880733} & SO(3)$\times$SO(3) & -- & 30.667 & 0.6812 & 1.4679 & $-0.3862$ & 0.4955 & U & U
\\
{\tt S0983994} & SO(3)$\times$U(1) & -- & 21.852 & 1.1077 & 1.6399 & $-0.2281$ & 0.8743 & U & U
\\
{\tt S1039230} & SO(3)$\times$SO(3) & -- & 19.392 & 1.3923 & 1.7320 & $-0.1361$ & 1.1507 & U & U
\\
{\tt S1043471} & -- & -- & 30.787 & 1.0184 & 1.7391 & $-0.3095$ & 0.9469 & U & U
\\
{\tt S1176725} & -- & -- & 47.660 &  1.0915 &  1.9612 & $-0.3557$ &  1.1707 & U & U 
\\
{\tt S1200000} & U(1)$\times$U(1) & $\mathcal{N}=1$ & 29.856 & 2.0096 & 2.0000 & +0.0034 & 1.4248 & S & S
\\
{\tt S1384096} & SO(3) & $\mathcal{N}=1$ & {\it 65.874} & 2.3216  & 2.3068 & +0.0049 & & S & S
\\
{\tt S1400000} & SO(3)$\times$SO(3) & -- & 39.973 & 2.0000  & 2.3333 & $-0.1133$ & 1.6645 & S & U
\\
{\tt S1424025} & SO(3) & -- & 56.525 & 2.0867 & 2.3734 & $-0.0960$ & 1.6232 & U & U
\\
{\tt S1600000} & $\IZ_3$ & -- &  93.255 &  1.7157 &  2.6667 &  $-0.3231$ &  $2.3148 $ & U & U 
\\
{\tt S1800000} & -- & -- & 107.46 &  1.5525 &  3.0000 & $-0.4861$ &  2.67807 & U & U
\\
{\tt S2095412} & -- & -- &   214.16 &  2.5031 &  3.4924 &  $-0.2867$ &  3.6521 & U & U 
\\
{\tt S2096313} & SO(3)$\times$U(1) & -- & 109.18 & 4.3673 & 3.4939 & $+0.2206$ & 4.3254 & U & S
\\
{\tt S2153574} & -- & -- & 210.73 &  2.2935 & 3.5893 &  $-0.1577$ &  3.8441 & U & U 
\\
{\tt S2443607} & SO(3) & -- & 203.21 & {\it 3.6949}  & 4.0727 & & 6.5629 & U & U
\\
{\tt S2702580} & $\IZ_3$ & -- & 338.08 &  3.1256 & 4.5043 & $ -0.3544 $ &  5.9984 & U & U 
\\
{\tt S3254576} & -- & --  & 511.14 & 2.4228 & 5.4243 &$ -0.7725 $ &  6.6393 & U & U 
\\
{\tt S3305153} & $\IZ_3$ & -- & 668.38 & 4.7754 &   5.5086& $ -0.1618 $ &  9.0309 & U & U
\\
{\tt S4599899} & -- & --  & 1957.8 & 6.7127 & 7.6665 & $-0.1780 $ & 13.081 & U & U\\
\noalign{\smallskip}
\bottomrule
\end{tabular}}
\caption{\label{tblpts} Numerical estimates for the stability parameters of some of critical points of gauged $\Neql8$  supergravity in four dimensions.   The quantity $\Theta_\text{avg}^{\le}$  is a lower bound on the average of $\Theta$. ``BF'' and ``BJ'' catalogue the Breitenlohner-Freedman and Brane-Jet (in)stability  with S=stable, and  U=unstable. The italicized entries are based on an approximation discussed in Appendix \ref{App:Uplifts}.}
\end{center}
\end{table}

Thanks to the development of new search algorithms \cite{Fischbacher:2008zu,Comsa:2019rcz},  we now know vastly more about the critical points of gauged supergravities.  In particular,  the details of 194 critical points of the potential of the maximal gauged supergravity in four dimensions are now known, either analytically or numerically, to a very high precision  \cite{Fischbacher:2011jx,Comsa:2019rcz, Bobev:2019dik}. For each such point, one can evaluate, explicitly,  the corresponding scalar fields and obtain the warp factor, $\Delta^{-3}$, from the uplift formula \eqref{Deltam3}. 

The results of those estimates for a sample of 26 points\footnote{Following the standard convention
\cite{Fischbacher:2011jx}, the points are labelled as $\tt Sn_1n_2n_3n_4n_5n_6n_7$, which encodes the first seven digits of the cosmological constant, $P_*={\tt n_1n_2.n_3n_4n_5n_6n_7}$.}
  are presented in Table~\ref{tblpts}.  Those points include all known supersymmetric points; the $SO(3) \times SO(3)$ invariant, non-supersymmetric point, $\tt S1400000$;  all points that are invariant under the triality symmetric $\rm SO(3)$ classified in  \cite{Bobev:2019dik}, and  some additional points with no continuous symmetry chosen to cover the whole range of the cosmological constant, from the first  such point, $\tt S1043471$, to the lowest lying known point, $\tt S4599899$.

For reasons outlined in Appendix \ref{App:Uplifts}, we have catalogued $\Delta^{-3}$ and   $P_*/6$ (as opposed to the square-roots of these quantities).  The quantity $\Theta_\text{avg}^{\le}$  is an easily computed lower bound on the average of $\Theta$.  See Appendix \ref{App:Uplifts} for its definition and details. The important point is that the average value of $\Theta$ is always positive, which means that smeared branes do not see brane-jet instabilities.  The last columns, ``BF'' and ``BJ'' catalogue the Breitenlohner-Freedman and Brane-Jet stability  with S=stable, and  U=unstable.  

All the supersymmetric points are BJ stable, as one would expect.  Also, as discussed in Sections \ref{ss:G2SO7} and \ref{ss:SU4},  highly symmetric BF unstable points can be BJ stable because the symmetry effectively averages the brane potential over large regions of the $S^7$.   Interestingly enough, we also find one point, $\tt S2096313$, that has relatively little symmetry and yet is BJ stable but BF unstable.  It would be interesting to understand why the brane probes do not detect the instability of this point.

The most significant result is, of course, $\tt S1400000$, which is BF stable but BJ unstable.  This result now means that every known non-supersymmetric AdS vacuum arising from critical points of gauged supergravity in four or five dimensions is unstable.  Thus the conjectured instability of non-supersymmetric AdS vacua is, once again, vindicated.

\section{Final comments}
\label{sec:Conclusions}
 
We have exhibited a new brane-jet instability that can arise in  AdS vacua of higher-dimensional theories in which the metric has a  non-trivial warp factor in front of the AdS factor.    While we have studied this in detail for AdS$_4$ vacua associated with M2 branes, there are obvious generalizations to other brane systems and AdS spaces in various dimensions. 

Describing the instability as a brane jet is intended to connote the fact that such instabilities are typically localized on the manifold that surround the brane system.  In the 26 examples we studied, we found  many examples of brane jets.  We also found that if a brane jet were present, then it was also quite weak.  Indeed, we found that if we averaged the potential of the brane probe over the sphere, then the result was always positive and the instability disappeared.  It would be extremely interesting to understand if this is true for all the other vacua of gauged maximal supergravity as it may have interesting consequences for the low-dimensional effective field theories.  In this context, it is interesting to note that, at least in massive IIA theories, there can be uniformly negative brane potentials \cite{Gaiotto:2009mv}.

It is also evident that, while brane-jet instabilities are a useful tool, they  have their limitations.  After all,  these instabilities only come through a single function, the warp factor, and this can be a blunt instrument. Indeed,  we have seen that BF unstable vacua can be BJ stable, and this is not simply an artifact of symmetry: one example of a BF-unstable/BJ-stable vacuum has merely an $SO(3) \times U(1)$ symmetry. 

The utility of brane-jet instabilities lies in their easy application:  they allow one to test a new class of intuitive and simple decay channels.  In particular, we have shown that there is a brane-jet instability  in the $SO(3) \times SO(3)$ invariant vacuum of $\Neql8$ gauged supergravity in four dimensions despite the perturbative stability within the $\Neql8$ gauged supergravity.  This vacuum is therefore unstable, as conjectured in \cite{Ooguri:2016pdq}.

Based on the fact that supergravity scalars in low dimensions encode Coulomb branches of the branes, we have argued that brane-jet instabilities are simply another way of approaching instabilities that appear in the complete scalar spectrum in the lower dimensional theory. Roughly, brane jets use delta-function probes while scalar modes use Fourier series.   In particular, gauged maximal supergravities are limited to ``lowest harmonics'' that may not resolve and detect a brane-jet instability.  To see brane-jet instabilities using scalars one may have to use higher harmonics and possibly non-trivial combinations of them.  Conversely, in the vacua that are BF-unstable but BJ-stable, one may have to evolve the BF unstable mode some distance before a brane-jet instability emerges.  

There have been some recent developments that may enable one to test these ideas.  For AdS$_4$ vacua arising from scalar vevs in  gauged $\Neql8$ supergravity, there are now methods to compute the spectra of all the eleven-dimensional Kaluza-Klein towers above these vacua  \cite{Malek:2019eaz}.  As the authors of \cite{Malek:2019eaz} remark, it would be very interesting to use their technology to look for instabilities in all the Kaluza-Klein towers above the $SO(3)\times SO(3)$ invariant vacuum.  Indeed, we understand that there will be a forthcoming paper \cite{Malek:2020mlk} that examines the perturbative stability of the $SO(3) \times SO(3)$ invariant vacuum.
 
There are  many interesting AdS vacua in other dimensions and there are obvious generalizations to D3 and M5 brane jets. There are also possible D1 or D5 brane jets in the AdS$_3$ vacua of six-dimensional supergravities.  Indeed, there are quite a few BF-stable, non-supersymmetric vacua in gauge supergravity in three-dimensions\footnote{Private communication from T.~Fischbacher and H.~Nicolai.}.  While a significant fraction of the three-dimensional gauged supergravities have unknown, or perhaps, non-existent higher dimensional analogues, it would be interesting to see if there were any BF-stable, non-supersymmetric vacua that do have higher-dimensional uplifts and then look for brane-jet instabilities.

\section*{Acknowledgments}
\vspace{-2mm}
We are very grateful to Cumrun Vafa for discussions about instabilities in AdS$_4$ vacua.  KP would like to thank  Bruno de Luca for a discussion and correspondence.  The work of IB  is supported by the ANR grant Black-dS-String ANR-16-CE31-0004-01 and by the John Templeton Foundation grant 61149.   The work of KP and NW was supported in part by the DOE grant DE-SC0011687 and IB and NW  were also supported by the ERC Grant 787320 - QBH Structure.  KP is grateful to the IPhT Saclay for hospitality during the initial stage of this project.


\vspace{1cm}

\appendix
\leftline{\LARGE \bf Appendices}

\section{Uplift formulae for the metric and the warp factor}
\label{App:Uplifts}

In this appendix we summarize briefly an explicit construction of the warp factor, $\Delta$, for any point on the scalar coset manifold, $\rm E_{7(7)}/SU(8)$, of the four-dimensional supergravity using standard scalar harmonics, $Y^A$, $A=1,\ldots,8$,  on $S^7$. Those harmonics can be viewed as the Cartesian coordinates of the ambient space, $\IR^8$, satisfying $Y^AY^A= m_7^{-2}$ when restricted to the sphere. 
 
We start with a   point on the scalar coset given by the scalar 56-bein \cite{Cremmer:1979up,deWit:1982bul},
\begin{equation}\label{56bein}
\mathcal{V}\eql  \left(\begin{matrix}
u_{ij}{}^{IJ} & v_{ijKL}\\ v^{kl IJ} & u^{kl}{}_{KL}
\end{matrix}\right)\eql \exp \left(\begin{matrix}
0 & -\coeff 1 4 \sqrt 2\,\phi_{ijkl}\\
 -\coeff 1 4 \sqrt 2\,\bar\phi^{ijkl} & 0
\end{matrix}\right) \in {\rm E_{7(7)}}\,,
\end{equation}
 in the symmetric gauge  and in the $\rm SU(8)$ basis. Its rotation to the  $\rm SL(8,\IR)$ basis is  \cite{Cremmer:1979up}\footnote{The $\rm SO(8)$ gamma matrices, $\Gamma_{AB}=-\Gamma_{BA}$, are defined as
\begin{equation*}
\Gamma_{ab}{}\eql \Gamma_{[a}\Gamma_{b]}\,,\qquad \Gamma_{a8}\eql-\Gamma_{8a}\eql -i\,\Gamma_a\,,\qquad \Gamma_{88}=0\,,
\end{equation*}
where $\Gamma_a$ are $\rm SO(7)$ gamma matrices satisfying $\Gamma_1\Gamma_2\ldots\Gamma_7\eql -i$. One can choose  $\Gamma_a$ to be pure imaginary and hermitian, hence antisymmetric. The $\rm SO(8)$ commutators are then
 $
[\Gamma_{AB},\Gamma_{CD}]\eql -2\delta_{AC}\Gamma_{BD}+\ldots\,.
$}
\begin{equation}\label{}
\begin{split}
U_{ij}{}^{AB} & \eql u_{ij}{}^{IJ}(\Gamma_{AB})^{IJ}\,,\qquad V_{ijAB} \eql v_{ijIJ}(\Gamma_{AB})^{IJ}\,,\\
U^{ij}{}_{AB} & \eql u^{ij}{}_{IJ}(\Gamma_{AB})^{IJ}\,,\qquad V^{ijAB} \eql v^{ijIJ}(\Gamma_{AB})^{IJ}\,.
\end{split}
\end{equation}
Define the matrix
\begin{equation}\label{}
M_{AB\,CD}\eql (U^{ij}{}_{AB}+V^{ijAB})(U_{ij}{}^{CD}+V_{ijCD})\,,
\end{equation}
which is both real and symmetric in $(AB,CD)$. 
The tensor
\begin{equation}\label{}
G^{AB}\eql M_{AC\,BD}Y^CY^D\,,
\end{equation}
in $\IR^8$, is then transverse to the sphere, $G^{AB}Y^B=0$. By extending the coordinates, $y^m$, $m=1,\ldots,7$, on $S^7$ to ``sperical coordinates'' on $\IR^8$ by a radial coordinate, $r$, we then have
\begin{equation}\label{Gup}
G^{mn}\eql \Delta^{-1}g^{mn}\,,\qquad G^{mr}\eql G^{rm}\eql G^{rr}\eql 0\,,
\end{equation}
where $g^{mn}$ is the uplift of the inverse metric \cite{deWit:1984nz}.\footnote{Perhaps the more familiar way of writing the same uplift formula  is
\begin{equation*}
\Delta^{-1}g^{mn}\eql {1\over 8} M_{AB\,CD}\mathcal{K}^{ABm}\mathcal{K}^{CDn}\,,\qquad \mathcal{K}^{ABm}\eql m_7\,\go^{mn}Y^{[A}\partial_n Y^{B]}\,,
\end{equation*}
where $\mathcal{K}^{ABm}$ are the $\rm SO(8)$ Killing vectors on $S^7$.
} 

For highly symmetric critical points, using coordinates, $y^m$,  such that  the isometries of the corresponding uplift   are manifest,  it is usually quite straightforward to calculate both the warp factor, $\Delta$, and the metric tensor, $g_{mn}$, starting from $G^{mn}$ in \eqref{Gup} (see, e.g.\ \cite{deWit:1984nz,Corrado:2001nv,Godazgar:2014eza}),
\begin{equation}\label{DDelta}
\Delta \eql \left[\det(G^{mn})\det(\go_{mn})\right]^{-1/9}\,,\qquad (g_{mn})\eql \Delta^{-1} (G^{mn})^{-1}\,.
\end{equation}
 However,  for points with   a large number of non-vanishing scalars, $\phi_{ijkl}$,  and the resulting low or no symmetry, this calculation quickly becomes quite involved. It is only recently that a more practical formula for the uplifted metric has been obtained in \cite{Varela:2015ywx,Kruger:2016agp} by exploiting a cubic invariant  of $\rm E_{7(7)}$. In terms of the ambient harmonics we are using here, it  works as follows.\footnote{Our discussion below is based on \cite{Kruger:2016agp}.}

Define two complex one-forms in $\IR^8$,
\begin{equation}\label{}
\begin{split}
\mathcal{A} _{ijkl} & ~\equiv~
  -{m_7\over 4}\,Y^{(A}dY^{B)}(U_{ij}{}^{AC}U_{kl}{}^{BC}-V_{ijAC}V_{klBC})\,,\\
\mathcal{B} _{ijkl} & ~\equiv~
  -{m_7\over 4}\,Y^{[A}dY^{B]}\,(U_{ij}{}^{AC}V_{klBC}-V_{ijAC}U_{kl}{}^{BC})\,,
\end{split}
\end{equation}
which are completely antisymmetric and (complex) self-dual in their $\rm SU(8)$ indices $[ijkl]$. Then the pull-back  of the symmetric tensor\footnote{As usual, the raised/lowered $\rm SU(8)$ indices denote complex conjugation.} 
\begin{equation}\label{}
G_{AB}dY^A dY^B~\equiv~{1\over 12}(\mathcal{A} _{ijkl}-\mathcal{B} _{ijkl})(\mathcal{A} ^{ijkl}-\mathcal{B} ^{ijkl})\,,
\end{equation}
onto $S^7$ is proportional to the uplifted metric,
\begin{equation}\label{Gdn}
G_{mn}\eql \Delta^{-2}\,g_{mn}\,.
\end{equation}
Moreover, from \eqref{Gup} and \eqref{Gdn}, we have \cite{Varela:2015ywx,Kruger:2016agp}
\begin{equation}\label{Deltam3}
\Delta^{-3}\eql {1\over 7}\,(\Delta^{-1}g^{mn})(\Delta^{-2}g_{mn})\eql {1\over 7}\,G^{AB}G_{AB}\,.
\end{equation}

An obvious advantage of the formulae \eqref{Gdn} and  \eqref{Deltam3} over \eqref{DDelta} is that one obtains the internal metric and the warp factor corresponding to a given scalar 56-bein \eqref{56bein} using only multiple summations. This procedure yields $\Delta^{-3}$ given by a homogenous polynomial of order four in the ambient scalar harmonics.

To find the maximum and/or minimum of $\Delta^{-3}$ on $S^7$ one can use any of the readily available numerical routines such as for instance $\tt FindMaximum[~\cdot~]$ and $\tt FindMinimum[~\cdot~]$ in Mathematica \cite{Mathematica}.
Then
\begin{equation}\label{}
\Theta_\text{min}\eql \sqrt{\Delta_\text{min}^{-3}}-\sqrt{P_*\over 6}\,.
\end{equation}
From the obvious identity
\begin{equation}\label{}
\begin{split}
\Delta^{-3}-{P_*\over 6} & \leq \left(\Delta^{-3/2}-\sqrt{P_*\over 6}\,\right)\left(\sqrt{\Delta^{-3}_\text{max}}+\sqrt{P_*\over 6}\,\right)\\
& \eql \Theta\,\left(\sqrt{\Delta^{-3}_\text{max}}+\sqrt{P_*\over 6}\,\right)\,,
\end{split}
\end{equation}
we get a lower bound,  $\Theta_\text{avg}^{\le}$, for the average force, $\Theta_\text{avg}$,
\begin{equation}\label{ThAVle}
\Theta_\text{avg}^{\le}\equiv {\text{vol}_{S^7}^{-1}\over \sqrt{\Delta^{-3}_\text{max}}+\sqrt{P_*/ 6}}\,\int _{S^7} \left(\Delta^{-3}-{P_*\over 6}\right)\,d\text{vol}_{S^7} \leq \Theta_\text{avg}\equiv {1\over \text{vol}_{S^7}}\int_{S^7}\,\Theta\,d\text{vol}_{S^7}\,,
\end{equation}
where $\text{Vol}_{S^7}={1\over 3}\pi^4$.

The integral on the left hand side  in \eqref{ThAVle} can be evaluated algebraically using the following overlap integrals for the scalar harmonics on the unit $S^7$,
\begin{equation}\label{}
\int_{S^7}\,Y^AY^BY^CY^D\,d\text{vol}_{S^7}\eql m_7^{-4}\,{\pi^4\over 240}(\delta^{AB}\delta^{CD}+\delta^{AC}\delta^{BD}+\delta^{AD}\delta^{BC})\,.
\end{equation}

The  numerical estimates based on the foregoing formulae for a sample of 26 critical points are listed in Table~\ref{tblpts}. The two italicized entries are approximate, where for the point  $\tt S1384096$, $\Delta_\text{max}^{-3}\geq 65.874$, while for $\tt S2443607$, $\Delta_\text{min}^{-3} \leq 3.6949$. Clearly, these bounds are sufficient for our analysis.  

\vfill\eject


\newpage
\begin{adjustwidth}{-1mm}{-1mm} 
\bibliographystyle{JHEP}      
\bibliography{stability}       

\providecommand{\href}[2]{#2}\begingroup\raggedright\begin{thebibliography}{10}

\bibitem{Breitenlohner:1982bm}
P.~Breitenlohner and D.~Z. Freedman, \emph{{Positive Energy in anti-De Sitter
  Backgrounds and Gauged Extended Supergravity}},
  \href{https://doi.org/10.1016/0370-2693(82)90643-8}{\emph{Phys. Lett.}
  {\bfseries 115B} (1982) 197}.

\bibitem{Breitenlohner:1982jf}
P.~Breitenlohner and D.~Z. Freedman, \emph{{Stability in Gauged Extended
  Supergravity}},
  \href{https://doi.org/10.1016/0003-4916(82)90116-6}{\emph{Annals Phys.}
  {\bfseries 144} (1982) 249}.

\bibitem{Gibbons:1983aq}
G.~W. Gibbons, C.~M. Hull and N.~P. Warner, \emph{{The Stability of Gauged
  Supergravity}},
  \href{https://doi.org/10.1016/0550-3213(83)90480-7}{\emph{Nucl. Phys.}
  {\bfseries B218} (1983) 173}.

\bibitem{Bobev:2011rv}
N.~Bobev, A.~Kundu, K.~Pilch and N.~P. Warner, \emph{{Minimal Holographic
  Superconductors from Maximal Supergravity}},
  \href{https://doi.org/10.1007/JHEP03(2012)064}{\emph{JHEP} {\bfseries 03}
  (2012) 064} [\href{https://arxiv.org/abs/1110.3454}{{\ttfamily 1110.3454}}].

\bibitem{ArkaniHamed:2006dz}
N.~Arkani-Hamed, L.~Motl, A.~Nicolis and C.~Vafa, \emph{{The String landscape,
  black holes and gravity as the weakest force}},
  \href{https://doi.org/10.1088/1126-6708/2007/06/060}{\emph{JHEP} {\bfseries
  06} (2007) 060} [\href{https://arxiv.org/abs/hep-th/0601001}{{\ttfamily
  hep-th/0601001}}].

\bibitem{Ooguri:2016pdq}
H.~Ooguri and C.~Vafa, \emph{{Non-supersymmetric AdS and the Swampland}},
  \href{https://doi.org/10.4310/ATMP.2017.v21.n7.a8}{\emph{Adv. Theor. Math.
  Phys.} {\bfseries 21} (2017) 1787}
  [\href{https://arxiv.org/abs/1610.01533}{{\ttfamily 1610.01533}}].

\bibitem{Warner:1983du}
N.~P. Warner, \emph{{Some Properties of the Scalar Potential in Gauged
  Supergravity Theories}},
  \href{https://doi.org/10.1016/0550-3213(84)90286-4}{\emph{Nucl. Phys.}
  {\bfseries B231} (1984) 250}.

\bibitem{Warner:1983vz}
N.~P. Warner, \emph{{Some New Extrema of the Scalar Potential of Gauged $N=8$
  Supergravity}},
  \href{https://doi.org/10.1016/0370-2693(83)90383-0}{\emph{Phys. Lett.}
  {\bfseries 128B} (1983) 169}.

\bibitem{Fischbacher:2009cj}
T.~Fischbacher, \emph{{Fourteen new stationary points in the scalar potential
  of SO(8)-gauged N=8, D=4 supergravity}},
  \href{https://doi.org/10.1007/JHEP09(2010)068}{\emph{JHEP} {\bfseries 09}
  (2010) 068} [\href{https://arxiv.org/abs/0912.1636}{{\ttfamily 0912.1636}}].

\bibitem{Fischbacher:2010ec}
T.~Fischbacher, K.~Pilch and N.~P. Warner, \emph{{New Supersymmetric and
  Stable, Non-Supersymmetric Phases in Supergravity and Holographic Field
  Theory}},  \href{https://arxiv.org/abs/1010.4910}{{\ttfamily 1010.4910}}.

\bibitem{Fischbacher:2011jx}
T.~Fischbacher, \emph{{The Encyclopedic Reference of Critical Points for
  SO(8)-Gauged N=8 Supergravity. Part 1: Cosmological Constants in the Range
  $-\Lambda/g^2 \in [6:14.7)$ }},
  \href{https://arxiv.org/abs/1109.1424}{{\ttfamily 1109.1424}}.

\bibitem{Borghese:2013dja}
A.~Borghese, A.~Guarino and D.~Roest, \emph{{Triality, Periodicity and
  Stability of SO(8) Gauged Supergravity}},
  \href{https://doi.org/10.1007/JHEP05(2013)107}{\emph{JHEP} {\bfseries 05}
  (2013) 107} [\href{https://arxiv.org/abs/1302.6057}{{\ttfamily 1302.6057}}].

\bibitem{Comsa:2019rcz}
I.~M. Comsa, M.~Firsching and T.~Fischbacher, \emph{{SO(8) Supergravity and the
  Magic of Machine Learning}},
  \href{https://doi.org/10.1007/JHEP08(2019)057}{\emph{JHEP} {\bfseries 08}
  (2019) 057} [\href{https://arxiv.org/abs/1906.00207}{{\ttfamily
  1906.00207}}].

\bibitem{Bobev:2019dik}
N.~Bobev, T.~Fischbacher and K.~Pilch, \emph{{Properties of the new $
  \mathcal{N} $ = 1 AdS$_{4}$ vacuum of maximal supergravity}},
  \href{https://doi.org/10.1007/JHEP01(2020)099}{\emph{JHEP} {\bfseries 01}
  (2020) 099} [\href{https://arxiv.org/abs/1909.10969}{{\ttfamily
  1909.10969}}].

\bibitem{Nicolai:1985hs}
H.~Nicolai and N.~P. Warner, \emph{{The SU(3) X U(1) Invariant Breaking of
  Gauged $N=8$ Supergravity}},
  \href{https://doi.org/10.1016/0550-3213(85)90643-1}{\emph{Nucl. Phys.}
  {\bfseries B259} (1985) 412}.

\bibitem{Bobev:2010ib}
N.~Bobev, N.~Halmagyi, K.~Pilch and N.~P. Warner, \emph{{Supergravity
  Instabilities of Non-Supersymmetric Quantum Critical Points}},
  \href{https://doi.org/10.1088/0264-9381/27/23/235013}{\emph{Class. Quant.
  Grav.} {\bfseries 27} (2010) 235013}
  [\href{https://arxiv.org/abs/1006.2546}{{\ttfamily 1006.2546}}].

\bibitem{Khavaev:1998fb}
A.~Khavaev, K.~Pilch and N.~P. Warner, \emph{{New vacua of gauged N=8
  supergravity in five-dimensions}},
  \href{https://doi.org/10.1016/S0370-2693(00)00795-4}{\emph{Phys. Lett.}
  {\bfseries B487} (2000) 14}
  [\href{https://arxiv.org/abs/hep-th/9812035}{{\ttfamily hep-th/9812035}}].

\bibitem{Krishnan:2020sfg}
C.~Krishnan, V.~Mohan and S.~Ray, \emph{{Machine learning ${\cal N}=8, D=5$
  gauged supergravity}},  \href{https://arxiv.org/abs/2002.12927}{{\ttfamily
  2002.12927}}.

\bibitem{Bobev:2020ttg}
N.~Bobev, T.~Fischbacher, F.~F. Gautason and K.~Pilch, \emph{{A Cornucopia of
  AdS$_5$ Vacua}},  \href{https://arxiv.org/abs/2003.03979}{{\ttfamily
  2003.03979}}.

\bibitem{Freedman:1999gp}
D.~Z. Freedman, S.~S. Gubser, K.~Pilch and N.~P. Warner, \emph{{Renormalization
  group flows from holography supersymmetry and a c theorem}},
  \href{https://doi.org/10.4310/ATMP.1999.v3.n2.a7}{\emph{Adv. Theor. Math.
  Phys.} {\bfseries 3} (1999) 363}
  [\href{https://arxiv.org/abs/hep-th/9904017}{{\ttfamily hep-th/9904017}}].

\bibitem{Freedman:1999dz}
D.~Z. Freedman, S.~S. Gubser, K.~Pilch and N.~P. Warner, ``{Private
  communication with \hbox{J.\ Distler and F.\ Zamora}}.'' Unpublished, 1999.

\bibitem{Pilch:1999misc}
K.~Pilch, ``{\it Notes on perturbative instability of the SO(5),
  SU(2)$\times$U(1)$\times$U(1), and SU(3) AdS$_5$ vacua}.'' Unpublished, 1999.

\bibitem{Distler:1999tr}
J.~Distler and F.~Zamora, \emph{{Chiral symmetry breaking in the AdS / CFT
  correspondence}},
  \href{https://doi.org/10.1088/1126-6708/2000/05/005}{\emph{JHEP} {\bfseries
  05} (2000) 005} [\href{https://arxiv.org/abs/hep-th/9911040}{{\ttfamily
  hep-th/9911040}}].

\bibitem{Girardello:1998pd}
L.~Girardello, M.~Petrini, M.~Porrati and A.~Zaffaroni, \emph{{Novel local CFT
  and exact results on perturbations of N=4 superYang Mills from AdS
  dynamics}}, \href{https://doi.org/10.1088/1126-6708/1998/12/022}{\emph{JHEP}
  {\bfseries 12} (1998) 022}
  [\href{https://arxiv.org/abs/hep-th/9810126}{{\ttfamily hep-th/9810126}}].

\bibitem{Boucher:1984yx}
W.~Boucher, \emph{{Positive Energy without Supersymmetry}},
  \href{https://doi.org/10.1016/0550-3213(84)90394-8}{\emph{Nucl. Phys.}
  {\bfseries B242} (1984) 282}.

\bibitem{Johnson:2000ic}
C.~V. Johnson, K.~J. Lovis and D.~C. Page, \emph{{Probing some N=1 AdS / CFT RG
  flows}}, \href{https://doi.org/10.1088/1126-6708/2001/05/036}{\emph{JHEP}
  {\bfseries 05} (2001) 036}
  [\href{https://arxiv.org/abs/hep-th/0011166}{{\ttfamily hep-th/0011166}}].

\bibitem{Corrado:2001nv}
R.~Corrado, K.~Pilch and N.~P. Warner, \emph{{An N=2 supersymmetric membrane
  flow}}, \href{https://doi.org/10.1016/S0550-3213(02)00134-7}{\emph{Nucl.
  Phys.} {\bfseries B629} (2002) 74}
  [\href{https://arxiv.org/abs/hep-th/0107220}{{\ttfamily hep-th/0107220}}].

\bibitem{Gaiotto:2009mv}
D.~Gaiotto and A.~Tomasiello, \emph{{The gauge dual of Romans mass}},
  \href{https://doi.org/10.1007/JHEP01(2010)015}{\emph{JHEP} {\bfseries 01}
  (2010) 015} [\href{https://arxiv.org/abs/0901.0969}{{\ttfamily 0901.0969}}].

\bibitem{Maldacena:1998uz}
J.~M. Maldacena, J.~Michelson and A.~Strominger, \emph{{Anti-de Sitter
  fragmentation}},
  \href{https://doi.org/10.1088/1126-6708/1999/02/011}{\emph{JHEP} {\bfseries
  02} (1999) 011} [\href{https://arxiv.org/abs/hep-th/9812073}{{\ttfamily
  hep-th/9812073}}].

\bibitem{Freedman:1999gk}
D.~Z. Freedman, S.~S. Gubser, K.~Pilch and N.~P. Warner, \emph{{Continuous
  distributions of D3-branes and gauged supergravity}},
  \href{https://doi.org/10.1088/1126-6708/2000/07/038}{\emph{JHEP} {\bfseries
  07} (2000) 038} [\href{https://arxiv.org/abs/hep-th/9906194}{{\ttfamily
  hep-th/9906194}}].

\bibitem{Carlisle:2003nd}
J.~E. Carlisle and C.~V. Johnson, \emph{{Holographic RG flows and universal
  structures on the Coulomb branch of N=2 supersymmetric large N gauge
  theory}}, \href{https://doi.org/10.1088/1126-6708/2003/07/039}{\emph{JHEP}
  {\bfseries 07} (2003) 039}
  [\href{https://arxiv.org/abs/hep-th/0306168}{{\ttfamily hep-th/0306168}}].

\bibitem{Gowdigere:2005wq}
C.~N. Gowdigere and N.~P. Warner, \emph{{Holographic Coulomb branch flows with
  N=1 supersymmetry}},
  \href{https://doi.org/10.1088/1126-6708/2006/03/049}{\emph{JHEP} {\bfseries
  03} (2006) 049} [\href{https://arxiv.org/abs/hep-th/0505019}{{\ttfamily
  hep-th/0505019}}].

\bibitem{Skenderis:2006di}
K.~Skenderis and M.~Taylor, \emph{{Holographic Coulomb branch vevs}},
  \href{https://doi.org/10.1088/1126-6708/2006/08/001}{\emph{JHEP} {\bfseries
  08} (2006) 001} [\href{https://arxiv.org/abs/hep-th/0604169}{{\ttfamily
  hep-th/0604169}}].

\bibitem{Witten:1981gj}
E.~Witten, \emph{{Instability of the Kaluza-Klein Vacuum}},
  \href{https://doi.org/10.1016/0550-3213(82)90007-4}{\emph{Nucl. Phys.}
  {\bfseries B195} (1982) 481}.

\bibitem{Horowitz:2007pr}
G.~T. Horowitz, J.~Orgera and J.~Polchinski, \emph{{Nonperturbative Instability
  of AdS(5) x S**5/Z(k)}},
  \href{https://doi.org/10.1103/PhysRevD.77.024004}{\emph{Phys. Rev.}
  {\bfseries D77} (2008) 024004}
  [\href{https://arxiv.org/abs/0709.4262}{{\ttfamily 0709.4262}}].

\bibitem{Narayan:2010em}
P.~Narayan and S.~P. Trivedi, \emph{{On The Stability Of Non-Supersymmetric AdS
  Vacua}}, \href{https://doi.org/10.1007/JHEP07(2010)089}{\emph{JHEP}
  {\bfseries 07} (2010) 089} [\href{https://arxiv.org/abs/1002.4498}{{\ttfamily
  1002.4498}}].

\bibitem{Ooguri:2017njy}
H.~Ooguri and L.~Spodyneiko, \emph{{New Kaluza-Klein instantons and the decay
  of AdS vacua}}, \href{https://doi.org/10.1103/PhysRevD.96.026016}{\emph{Phys.
  Rev.} {\bfseries D96} (2017) 026016}
  [\href{https://arxiv.org/abs/1703.03105}{{\ttfamily 1703.03105}}].

\bibitem{Apruzzi:2019ecr}
F.~Apruzzi, G.~Bruno De~Luca, A.~Gnecchi, G.~Lo~Monaco and A.~Tomasiello,
  \emph{{On AdS$_7$ stability}},
  \href{https://arxiv.org/abs/1912.13491}{{\ttfamily 1912.13491}}.

\bibitem{Borghese:2012qm}
A.~Borghese, A.~Guarino and D.~Roest, \emph{{All $G_2$ invariant critical
  points of maximal supergravity}},
  \href{https://doi.org/10.1007/JHEP12(2012)108}{\emph{JHEP} {\bfseries 12}
  (2012) 108} [\href{https://arxiv.org/abs/1209.3003}{{\ttfamily 1209.3003}}].

\bibitem{Guarino:2015vca}
A.~Guarino and O.~Varela, \emph{{Consistent $ \mathcal{N}=8 $ truncation of
  massive IIA on S$^{6}$}},
  \href{https://doi.org/10.1007/JHEP12(2015)020}{\emph{JHEP} {\bfseries 12}
  (2015) 020} [\href{https://arxiv.org/abs/1509.02526}{{\ttfamily
  1509.02526}}].

\bibitem{DallAgata:2011aa}
G.~Dall'Agata and G.~Inverso, \emph{{On the Vacua of N = 8 Gauged Supergravity
  in 4 Dimensions}},
  \href{https://doi.org/10.1016/j.nuclphysb.2012.01.023}{\emph{Nucl. Phys. B}
  {\bfseries 859} (2012) 70} [\href{https://arxiv.org/abs/1112.3345}{{\ttfamily
  1112.3345}}].

\bibitem{DallAgata:2014tph}
G.~Dall'Agata, G.~Inverso and A.~Marrani, \emph{{Symplectic Deformations of
  Gauged Maximal Supergravity}},
  \href{https://doi.org/10.1007/JHEP07(2014)133}{\emph{JHEP} {\bfseries 07}
  (2014) 133} [\href{https://arxiv.org/abs/1405.2437}{{\ttfamily 1405.2437}}].

\bibitem{Guarino:2015qaa}
A.~Guarino and O.~Varela, \emph{{Dyonic ISO(7) supergravity and the duality
  hierarchy}}, \href{https://doi.org/10.1007/JHEP02(2016)079}{\emph{JHEP}
  {\bfseries 02} (2016) 079}
  [\href{https://arxiv.org/abs/1508.04432}{{\ttfamily 1508.04432}}].

\bibitem{Danielsson:2017max}
U.~H. Danielsson, G.~Dibitetto and S.~C. Vargas, \emph{{A swamp of non-SUSY
  vacua}}, \href{https://doi.org/10.1007/JHEP11(2017)152}{\emph{JHEP}
  {\bfseries 11} (2017) 152}
  [\href{https://arxiv.org/abs/1708.03293}{{\ttfamily 1708.03293}}].

\bibitem{deWit:1984nz}
B.~de~Wit, H.~Nicolai and N.~P. Warner, \emph{{The Embedding of Gauged $N=8$
  Supergravity Into $d=11$ Supergravity}},
  \href{https://doi.org/10.1016/0550-3213(85)90128-2}{\emph{Nucl. Phys.}
  {\bfseries B255} (1985) 29}.

\bibitem{Nicolai:2011cy}
H.~Nicolai and K.~Pilch, \emph{{Consistent Truncation of d = 11 Supergravity on
  AdS$_4 \times S^7$}},
  \href{https://doi.org/10.1007/JHEP03(2012)099}{\emph{JHEP} {\bfseries 03}
  (2012) 099} [\href{https://arxiv.org/abs/1112.6131}{{\ttfamily 1112.6131}}].

\bibitem{Godazgar:2013nma}
H.~Godazgar, M.~Godazgar and H.~Nicolai, \emph{{Testing the non-linear flux
  ansatz for maximal supergravity}},
  \href{https://doi.org/10.1103/PhysRevD.87.085038}{\emph{Phys. Rev.}
  {\bfseries D87} (2013) 085038}
  [\href{https://arxiv.org/abs/1303.1013}{{\ttfamily 1303.1013}}].

\bibitem{Kruger:2016agp}
O.~Kruger, \emph{{Non-linear uplift Ansatze for the internal metric and the
  four-form field-strength of maximal supergravity}},
  \href{https://doi.org/10.1007/JHEP05(2016)145}{\emph{JHEP} {\bfseries 05}
  (2016) 145} [\href{https://arxiv.org/abs/1602.03327}{{\ttfamily
  1602.03327}}].

\bibitem{Kruger:2016yin}
O.~Kruger, \emph{{The embedding of gauged N = 8 supergravity into 11
  dimensions}}, Ph.D. thesis, Humboldt U., Berlin, 2016.
\newblock 10.18452/17662.

\bibitem{Ahn:2000aq}
C.-h. Ahn and J.~Paeng, \emph{{Three-dimensional SCFTs, supersymmetric domain
  wall and renormalization group flow}},
  \href{https://doi.org/10.1016/S0550-3213(00)00687-8}{\emph{Nucl. Phys.}
  {\bfseries B595} (2001) 119}
  [\href{https://arxiv.org/abs/hep-th/0008065}{{\ttfamily hep-th/0008065}}].

\bibitem{Ahn:2000mf}
C.-h. Ahn and K.~Woo, \emph{{Supersymmetric domain wall and RG flow from
  4-dimensional gauged N=8 supergravity}},
  \href{https://doi.org/10.1016/S0550-3213(01)00008-6}{\emph{Nucl. Phys.}
  {\bfseries B599} (2001) 83}
  [\href{https://arxiv.org/abs/hep-th/0011121}{{\ttfamily hep-th/0011121}}].

\bibitem{Klebanov:2008vq}
I.~Klebanov, T.~Klose and A.~Murugan, \emph{{AdS(4)/CFT(3) Squashed, Stretched
  and Warped}},
  \href{https://doi.org/10.1088/1126-6708/2009/03/140}{\emph{JHEP} {\bfseries
  03} (2009) 140} [\href{https://arxiv.org/abs/0809.3773}{{\ttfamily
  0809.3773}}].

\bibitem{Pope:1984bd}
C.~N. Pope and N.~P. Warner, \emph{{An SU(4) Invariant Compactification of
  $d=11$ Supergravity on a Stretched Seven Sphere}},
  \href{https://doi.org/10.1016/0370-2693(85)90992-X}{\emph{Phys. Lett.}
  {\bfseries 150B} (1985) 352}.

\bibitem{Godazgar:2014eza}
H.~Godazgar, M.~Godazgar, O.~Kruger, H.~Nicolai and K.~Pilch, \emph{{An
  SO(3)$\times$SO(3) invariant solution of $D=11$ supergravity}},
  \href{https://doi.org/10.1007/JHEP01(2015)056}{\emph{JHEP} {\bfseries 01}
  (2015) 056} [\href{https://arxiv.org/abs/1410.5090}{{\ttfamily 1410.5090}}].

\bibitem{Fischbacher:2008zu}
T.~Fischbacher, \emph{{The Many vacua of gauged extended supergravities}},
  \href{https://doi.org/10.1007/s10714-008-0736-z}{\emph{Gen. Rel. Grav.}
  {\bfseries 41} (2009) 315} [\href{https://arxiv.org/abs/0811.1915}{{\ttfamily
  0811.1915}}].

\bibitem{Malek:2019eaz}
E.~Malek and H.~Samtleben, \emph{{A Kaluza-Klein Spectrometer for
  Supergravity}},  \href{https://arxiv.org/abs/1911.12640}{{\ttfamily
  1911.12640}}.

\bibitem{Malek:2020mlk}
E.~Malek, H.~Nicolai and H.~Samtleben, \emph{{Tachyonic Kaluza-Klein modes and
  the AdS swampland conjecture}},
  \href{https://arxiv.org/abs/2005.07713}{{\ttfamily 2005.07713}}.

\bibitem{Cremmer:1979up}
E.~Cremmer and B.~Julia, \emph{{The SO(8) Supergravity}},
  \href{https://doi.org/10.1016/0550-3213(79)90331-6}{\emph{Nucl. Phys.}
  {\bfseries B159} (1979) 141}.

\bibitem{deWit:1982bul}
B.~de~Wit and H.~Nicolai, \emph{{N=8 Supergravity}},
  \href{https://doi.org/10.1016/0550-3213(82)90120-1}{\emph{Nucl. Phys.}
  {\bfseries B208} (1982) 323}.

\bibitem{Varela:2015ywx}
O.~Varela, \emph{{Complete $D=11$ embedding of SO(8) supergravity}},
  \href{https://doi.org/10.1103/PhysRevD.97.045010}{\emph{Phys. Rev.}
  {\bfseries D97} (2018) 045010}
  [\href{https://arxiv.org/abs/1512.04943}{{\ttfamily 1512.04943}}].

\bibitem{Mathematica}
.~Wolfram Research~Inc., ``Mathematica, version 12.0.''

\end{thebibliography}\endgroup
\end{adjustwidth}

\end{document}